\documentclass{article}

\usepackage{arxiv}

\usepackage[utf8]{inputenc} 
\usepackage[T1]{fontenc}    
\usepackage{hyperref}       
\usepackage{url}            
\usepackage{booktabs}       
\usepackage{amsfonts}       
\usepackage{nicefrac}       
\usepackage{microtype}      
\usepackage{lipsum}		
\usepackage{longtable}
\usepackage{pdflscape}
\usepackage{array}
\usepackage{graphicx}
\usepackage{natbib}
\usepackage{doi}
\usepackage{longtable}
\usepackage{geometry}

\title{The Impact of Large Language Multi-Modal Models on the Future Job Market}

\author{ \href{https://orcid.org/0000-0003-4170-1295}{\includegraphics[scale=0.06]{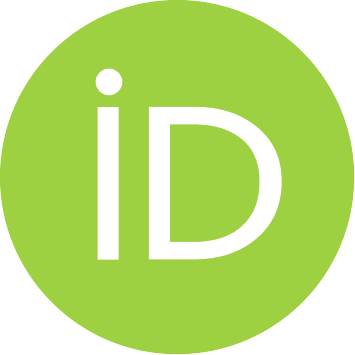}\hspace{1mm}Tarry Singh}\thanks{The author has co-written this paper with extreme haste since jobs are at massive risk and meanwhile he's preparing to run for the hills in case jobless zombies come for him.} \\
	Department of AI Research\\
	deepkapha AI Research\\
	9401 GE, Assen \\
        The Netherlands \\
	\texttt{tarry.singh@deepkapha.com} \\
}

\renewcommand{\shorttitle}{\ LLMs and Future of Jobs according to GPT-4}

\hypersetup{
pdftitle={A template for the arxiv style},
pdfsubject={q-bio.NC, q-bio.QM},
pdfauthor={David S.~Hippocampus, Elias D.~Striatum},
pdfkeywords={First keyword, Second keyword, More},
}

\begin{document}
\maketitle

\begin{abstract}
The rapid advancements in artificial intelligence, particularly in large language multi-modal models like GPT-4, have raised concerns about the potential displacement of human workers in various industries. This position paper aims to analyze the current state of job replacement by AI models and explores potential implications and strategies for a balanced coexistence between AI and human workers.
\end{abstract}

\keywords{LLM (Large Language Models) \and GPT-4 \and Future of Jobs \and Employment}

\section{Introduction}
The rapid advancement of AI and machine learning technology has resulted in a growing concern regarding job displacement in various industries. Large language models such as GPT-4 have shown impressive abilities in generating text, language translation, and even creating music and art. The applications of GPT-4 and similar models are numerous, ranging from chatbots and virtual assistants to content creation and data analysis. As these models become more sophisticated, they are bound to have a significant impact on the job market, especially in industries where automation can replace repetitive and monotonous tasks.

One of the sectors that are likely to experience the most significant impact is the content creation industry. The widespread use of GPT-4 and other large language models in generating text has raised concerns about the displacement of content writers and journalists. With the ability to generate high-quality content within minutes, these models could potentially replace human writers in the creation of news articles, reports, and even creative writing. The efficiency and accuracy of these models make them an attractive alternative to traditional content creation methods.

Another sector that is likely to experience significant job displacement is data processing. With the ability to analyze vast amounts of data within a fraction of the time it would take a human, GPT-4 could replace data analysts, who perform tasks such as data entry and data cleaning. Moreover, the model's ability to recognize patterns and make predictions could potentially eliminate the need for human analysts altogether. This could lead to a reduction in the demand for workers with these skills, which could potentially cause an oversupply of labor in the market.

However, it is important to note that while AI models such as GPT-4 have the potential to replace certain jobs, they can also create new opportunities for human workers. For example, as these models become more prevalent, the demand for AI engineers, data scientists, and machine learning experts is likely to increase. These highly skilled workers are necessary for the development and maintenance of these models, and their demand is expected to continue to grow as AI technology becomes more sophisticated.

Furthermore, as AI and machine learning become more integrated into the workplace, it is essential that workers develop skills that complement these technologies. Workers who possess skills such as critical thinking, creativity, and problem-solving are likely to remain valuable in the workforce, as these skills cannot be replicated by machines. In addition, the collaboration between humans and AI can result in more efficient and effective work processes, leading to increased productivity and better outcomes.

\subsection{Blue Collar Workforce}
The blue-collar workforce, traditionally associated with manual labor and manufacturing, has experienced significant changes over the past few decades, largely due to advances in technology and automation. LLMs like GPT-4 have accelerated this shift by automating jobs that require data analysis, pattern recognition, and even decision-making.

For instance, assembly line workers are being replaced by robots that can carry out tasks more efficiently, quickly, and safely. Likewise, the transportation industry is on the cusp of a major transformation with the advent of autonomous vehicles. Truck drivers, delivery personnel, and taxi operators face an uncertain future as companies invest in self-driving technology to reduce costs and increase productivity.

However, it is essential to note that LLM-driven automation will not eliminate all blue-collar jobs. Instead, it will likely shift the nature of these roles and create new opportunities for skilled workers. For example, technicians will be needed to maintain, repair, and upgrade automated systems. Furthermore, human workers will continue to be valuable in roles that require empathy, creativity, and critical thinking – skills that are difficult for AI to replicate.

\subsection{White Collar Workforce}
The impact of LLMs on white-collar jobs has been profound, as these models excel at automating tasks related to information processing, analysis, and communication. Legal professionals, financial analysts, customer service representatives, and even journalists have seen their roles disrupted by GPT-4 and other LLMs.

Legal professionals, for instance, are facing increasing competition from AI-driven services that can perform tasks like document review, contract analysis, and even legal research. Similarly, financial analysts are witnessing the rise of AI algorithms capable of processing vast amounts of data to make investment decisions or predict market trends.

In the realm of customer service, AI-powered chatbots are rapidly replacing human agents in handling routine inquiries, complaints, and troubleshooting issues. This allows companies to reduce costs while providing 24/7 support to customers. Lastly, journalism has also been affected by AI-generated content that can produce news articles or generate summaries with minimal human intervention.

Despite the growing influence of LLMs, white-collar professionals can adapt to the changing landscape by focusing on tasks that require unique human traits such as creativity, critical thinking, emotional intelligence, and interpersonal skills. These skills are difficult to replicate using AI models, and they will continue to be in demand in a world where automation and AI models are increasingly prevalent. Therefore, white-collar professionals should focus on developing these skills to complement AI technology and position themselves for new opportunities that arise in the future. In addition, professionals should stay up-to-date with advancements in AI and other emerging technologies to identify potential areas for collaboration and growth. This approach will allow white-collar professionals to thrive in a future where AI and human workers collaborate to create better outcomes for society.

\section{Job Replacement by AI Models - Serious Threat or A Bluff?}
As AI models such as GPT-4 become more sophisticated, the potential for job replacement becomes more significant. While the impact may vary across industries, it is essential to recognize that automation and AI have the potential to displace workers in many sectors, from manufacturing to service industries. The evolution of AI has been rapid, and it is expected that it will continue to grow in the coming years, increasing the potential for job displacement across various sectors.

The threat to white-collar jobs is becoming more apparent as AI models become better at performing tasks that were once thought to require human intelligence. For instance, GPT-4 can generate automated responses to customer queries, making it easier for customer support teams to manage their workload. In the legal sector, AI-powered machines can review contracts and legal documents, eliminating the need for paralegal workers. These examples illustrate how the growth of AI models is not only limited to blue-collar jobs but also white-collar jobs.

However, the potential for job displacement by AI models is not absolute. While AI is exceptional at performing repetitive tasks, it lacks certain human traits that are necessary for many jobs. Tasks that require creativity, empathy, and human judgment are less likely to be replaced by machines. In addition, human workers are essential for tasks that require face-to-face interaction, such as in healthcare and education. As a result, the role of human workers in the workforce is likely to evolve, rather than disappear altogether.

Furthermore, the growth of AI technology is expected to create new job opportunities. While some jobs may be replaced by AI, new roles will emerge that require skills that complement AI technology. For instance, AI technology is generating vast amounts of data that need to be analyzed, creating new roles in data science and analysis. In addition, the development and maintenance of AI models require skilled professionals such as AI engineers, software developers, and machine learning experts.

\section{Opportunities for Collaboration}
The concept of collaboration between AI models and human workers is gaining significant attention in the business world. While the growth of AI models has the potential to displace certain jobs, it is also creating opportunities for new types of jobs that require collaboration between humans and machines. This collaboration can lead to the creation of innovative solutions and a more efficient workforce.

One significant benefit of collaboration between humans and AI models is increased productivity. AI models are exceptional at performing repetitive tasks with a high degree of accuracy, while humans are better at performing complex tasks that require creativity and judgment. By combining the strengths of both AI and humans, productivity can increase significantly, leading to more efficient work processes and better outcomes.

In addition, collaboration between AI and human workers can lead to the creation of new job opportunities. For instance, as AI technology becomes more prevalent, there will be a greater need for professionals such as AI engineers, data scientists, and machine learning experts. These roles require high-level technical skills that are not easily replaceable by machines.

Moreover, AI technology can also lead to the creation of new industries that leverage the strengths of both AI and human workers. For instance, the healthcare industry is already experiencing the benefits of collaboration between AI and human workers. AI models are being used to analyze medical data and generate insights that can help doctors make more informed decisions. This collaboration can lead to the creation of new roles in the healthcare industry, such as medical data analysts and AI-assisted medical practitioners.

Furthermore, collaboration between humans and AI can lead to more creative solutions to complex problems. By leveraging AI models to automate repetitive tasks, human workers can focus on more complex tasks that require critical thinking and creativity. In this way, AI models can act as a tool to augment human capabilities, leading to more innovative solutions.

\section{Workforce Adaptability and Skills Development}
As AI models continue to advance, it is crucial for the workforce to adapt and develop new skills to remain relevant in an evolving job market. Emphasizing lifelong learning, reskilling, and upskilling can help workers remain competitive and secure their positions in an AI-driven world. Governments, educational institutions, and corporations must collaborate to create programs and opportunities that facilitate workforce transition and promote a culture of adaptability.

The advent of AI models such as GPT-4 has disrupted the traditional job market, creating the potential for job displacement across various industries. The impact of AI on the job market is not limited to just blue-collar jobs, but also white-collar jobs, which could face a significant risk of displacement. As AI technology continues to evolve, it is essential for the workforce to adapt and develop new skills to remain relevant in an evolving job market.

One of the most important ways for workers to adapt to the changing landscape is to emphasize lifelong learning. As technology evolves at an unprecedented rate, workers must be prepared to learn and adapt to new skills and tools throughout their careers. This requires a cultural shift that encourages and supports continuous learning and upskilling. Governments, educational institutions, and corporations must collaborate to create programs and opportunities that facilitate workforce transition and promote a culture of adaptability.

In addition to emphasizing lifelong learning, workers must also focus on reskilling and upskilling to remain competitive. As AI technology replaces certain jobs, it is essential for workers to acquire new skills and transition to new roles. This requires access to training programs and resources that enable workers to acquire the necessary skills to compete in an AI-driven world. In addition, workers must be willing to invest time and effort in learning new skills and adapting to new technologies.

Governments have an important role to play in promoting workforce adaptability and skills development. They can create policies and programs that incentivize employers to invest in employee training and development. For instance, governments can offer tax credits or subsidies to employers who invest in reskilling and upskilling their workers. They can also provide funding for training programs and apprenticeships to help workers acquire new skills and transition to new roles.

Educational institutions also have a crucial role to play in promoting workforce adaptability and skills development. They can offer courses and training programs that equip workers with the necessary skills to compete in an AI-driven world. In addition, they can partner with businesses and industry leaders to create training programs that align with industry needs and provide students with practical, hands-on experience.

Corporations also have a responsibility to invest in workforce adaptability and skills development. They can provide training programs and resources that enable workers to acquire new skills and transition to new roles. In addition, they can create a culture that emphasizes lifelong learning and encourages workers to acquire new skills and stay up-to-date with emerging technologies.

\section{Regulation and Policy}
The implementation of AI in the job market raises questions about ethical and legal considerations. Governments and regulatory bodies must establish policies that protect workers' rights, ensure fair competition, and prevent misuse of AI technologies. Additionally, public and private sectors should work together to address potential issues such as data privacy, AI bias, and the digital divide, which could exacerbate social inequality.

The rapid development of AI models has led to growing concerns about ethical and legal considerations surrounding the use of AI in the job market. Governments and regulatory bodies have an important role to play in establishing policies that protect workers' rights, ensure fair competition, and prevent the misuse of AI technologies.

The European Union (EU) has taken a significant step in regulating the use of AI with the recently introduced EU AI Law. The law aims to ensure that AI is used in a safe and ethical manner, protecting fundamental rights and European values. It includes provisions for the safety, transparency, and accountability of AI systems and the creation of a regulatory framework for high-risk AI applications. The EU AI Law is a significant step towards ensuring the responsible use of AI in the job market.

Data privacy is a crucial issue surrounding the use of AI in the job market. The General Data Protection Regulation (GDPR) is an EU regulation that governs the use of personal data and aims to protect individuals' privacy rights. The use of AI models in the job market requires the collection and processing of personal data, which must be done in compliance with GDPR regulations. It is essential for governments and businesses to ensure that AI models are developed and used in a manner that respects individuals' privacy rights.

In addition to privacy concerns, AI models must be developed and used in a manner that protects worker safety, security, and well-being. This includes ensuring that AI models are free from bias and do not discriminate against individuals based on factors such as age, gender, race, or ethnicity. AI models must also be designed to prioritize worker safety, whether in manufacturing or other industries where worker safety is a top priority.

Moreover, it is essential to consider the potential impact of AI on social inequality. The digital divide, or the gap between those who have access to technology and those who do not, is a significant issue that must be addressed. It is important for policymakers and businesses to work together to ensure that the benefits of AI are distributed equitably and that vulnerable populations are not left behind.

Worker protection is also a crucial issue that must be addressed in the development and use of AI models. Governments and businesses must establish policies that protect workers' rights, including the right to fair pay, reasonable working hours, and safe working conditions. Additionally, workers must be provided with opportunities for training and upskilling to remain competitive in an AI-driven world.

\section{Conclusions}
The potential for job replacement by large language multi-modal models like GPT-4 is an undeniable reality. However, the focus should be on fostering a symbiotic relationship between AI and human workers, enabling collaboration, adaptability, and skills development. Through effective regulation, policy-making, and education initiatives, the workforce can navigate the challenges posed by AI and leverage the technology's benefits to create a more prosperous and innovative future.
\begin{enumerate} 
\item  Collaboration between AI and human workers can lead to increased productivity, new job opportunities, and growth in industries that leverage the strengths of both AI and human workers.
\item Both white-collar and blue-collar jobs are at risk of displacement due to the increasing capabilities of AI and large language models such as GPT-4, but recognizing the potential for new job opportunities and the need for workforce adaptability is important.
\item Human workers will continue to be essential for tasks that require human judgment, creativity, and empathy, but the evolution of AI technology will create new job opportunities that require skills that complement AI technology.
\item Governments and regulatory bodies must establish policies that protect workers' rights, ensure fair competition, and prevent the misuse of AI technologies, while addressing potential issues such as data privacy, AI bias, and the digital divide that could exacerbate social inequality.
\item The future of work is likely to involve a hybrid approach where AI technology is used to automate mundane tasks, allowing human workers to focus on more complex and creative tasks, and collaboration between AI and human workers is crucial in creating a future where these technologies can be leveraged to their full potential.

\end{enumerate}

\appendix
\section{Appendix}

Below is a table generated by our colleague J'ai pété fort (Also pronounced GPT-4) with 50 jobs that GPT-4 could potentially replace, sorted in descending order of the percentage probability of replacement. Please note that the percentages are rough estimates and not absolute predictions. The average median salaries are approximate figures based on data available up to 2021 and may not accurately represent current salary levels.

\subsection{Fifty (50) Jobs at risk of being disrupted or replaced by GPT-4}
\begin{landscape}
\begin{small} 
\begin{longtable}{c>{\raggedright\arraybackslash}m{2.5cm}c>{\centering\arraybackslash}m{2.2cm}l>{\raggedright\arraybackslash}m{4cm}}
\caption{Jobs that could potentially be displaced or replaced by GPT-4} \\
\toprule
Number & Job & Median Salary (USD) & \%age Probability & Human Trait Replaced & Upskilling Advice \\
\midrule
\endhead
\bottomrule
\endfoot
1 & Claims Adjuster & 66,000 & 95\% & Insurance, Investigation & Learn fraud detection \\
2 & Stock Trader & 64,000 & 95\% & Trading, Financial Analysis & Gain expertise in niche markets \\
3 & Bookkeeper & 41,000 & 95\% & Accounting, Organization & Learn financial analysis \\
4 & Accountant & 71,000 & 95\% & Financial Analysis, Reporting & Specialize in tax or forensic accounting \\
5 & Data Analyst & 62,000 & 90\% & Data Analysis, Reporting & Learn machine learning, data science \\
6 & Medical Transcriptionist & 34,000 & 90\% & Typing, Medical Knowledge & Train in medical coding \\
7 & HR Coordinator & 47,000 & 90\% & Recruitment, Administration & Develop talent management skills \\
8 & Paralegal & 52,000 & 90\% & Legal Research, Document Preparation & Specialize in complex legal areas \\
9 & Personal Finance Advisor & 90,000 & 85\% & Financial Analysis, Planning & Focus on comprehensive financial planning \\
10 & Research Assistant & 32,000 & 85\% & Research, Data Analysis & Specialize in advanced research methods \\
11 & Proofreader & 37,000 & 80\% & Editing, Attention to Detail & Learn technical writing, copyediting \\
12 & Copywriter & 58,000 & 75\% & Writing, Marketing & Diversify into digital marketing \\
13 & Content Writer & 49,000 & 70\% & Research, Writing & Develop SEO, content strategy skills \\
14 & Online Tutor & 36,000 & 65\% & Education, Communication & Focus on personalized learning, soft skills \\
15 & Telemarketer & 27,000 & 60\% & Sales, Communication & Learn digital marketing, CRM skills \\
16 & Social Media Manager & 51,000 & 55\% & Content Management, Marketing & Learn video production, analytics \\
17 & Receptionist & 30,000 & 50\% & Communication, Organization & Acquire office management skills \\
18 & Virtual Assistant & 40,000 & 45\% & Organization, Task Management & Specialize in project management \\
19 & Customer Service Representative & 35,000 & 40\% & Communication, Problem Solving & Develop conflict resolution skills \\
20 & Data Entry Clerk & 31,000 & 35\% & Typing, Organization & Learn data visualization, analytics \\
21 & Translator & 51,000 & 30\% & Language, Translation & Specialize in legal, medical translation \\
22 & Editor & 61,000 & 25\% & Proofreading, Editing & Gain subject matter expertise \\
23 & Journalist & 45,000 & 20\% & Research, Writing & Learn multimedia journalism, data journalism \\
24 & Graphic Designer & 52,000 & 15\% & Creativity, Design & Master UX/UI design, 3D design \\
25 & Tour Guide & 27,000 & 10\% & Communication, Navigation & Develop local expertise, storytelling skills \\
26 & Software Developer & 107,000 & 10\% & Coding, Problem Solving & Learn AI programming, cybersecurity \\
27 & Public Relations Specialist & 62,000 & 10\% & Communication, Media Relations & Master crisis management, branding \\
28 & Digital Marketing Specialist & 57,000 & 10\% & Marketing, Analytics & Learn marketing automation, personalization \\
29 & Technical Writer & 74,000 & 10\% & Writing, Technical Knowledge & Specialize in high-demand industries \\
30 & Sales Representative & 52,000 & 10\% & Sales, Communication & Develop consultative selling, negotiation skills \\
31 & Event Planner & 50,000 & 10\% & Organization, Coordination & Learn virtual event planning, marketing \\
32 & Real Estate Agent & 50,000 & 10\% & Sales, Property Knowledge & Focus on niche markets, property management \\
33 & Photographer & 42,000 & 10\% & Creativity, Technical Skill & Master drone, 360-degree photography \\
34 & Occupational Therapist & 84,000 & 5\% & Healthcare, Rehabilitation & Specialize in geriatrics, mental health \\
35 & Architect & 80,000 & 5\% & Design, Technical Knowledge & Focus on sustainable, smart design \\
36 & Urban Planner & 74,000 & 5\% & Planning, Policy Analysis & Gain expertise in smart city planning \\
37 & Nurse & 75,000 & 5\% & Healthcare, Patient Care & Pursue specialization, advanced certifications \\
38 & Speech-Language Pathologist & 79,000 & 5\% & Healthcare, Communication & Specialize in teletherapy, autism \\
39 & Marriage and Family Therapist & 51,000 & 5\% & Counseling, Communication & Learn teletherapy, group therapy \\
40 & Dietitian & 63,000 & 5\% & Nutrition, Healthcare & Develop skills in telehealth, weight management \\
41 & Physician Assistant & 112,000 & 5\% & Healthcare, Diagnosis & Focus on rural healthcare, telemedicine \\
42 & Psychologist & 80,000 & 5\% & Counseling, Research & Specialize in trauma, addiction recovery \\
43 & Social Worker & 51,000 & 5\% & Counseling, Advocacy & Gain expertise in crisis intervention, policy \\
44 & Dentist & 159,000 & 5\% & Healthcare, Oral Care & Learn cosmetic dentistry, orthodontics \\
45 & Veterinarian & 96,000 & 5\% & Healthcare, Animal Care & Specialize in exotic pets, emergency care \\
46 & Physical Therapist & 89,000 & 5\% & Healthcare, Rehabilitation & Focus on telehealth, sports therapy \\
47 & Pharmacist & 128,000 & 5\% & Healthcare, Medication & Learn pharmaceutical care management \\
48 & Audiologist & 77,000 & 5\% & Healthcare, Hearing & Develop skills in cochlear implants, research \\
49 & Optometrist & 111,000 & 5\% & Healthcare, Vision & Master specialty contact lenses, low vision \\
50 & Surgeon & 409,000 & 5\% & Healthcare, Surgery & Specialize in minimally invasive techniques \\
\end{longtable}
\end{small}
\end{landscape}

\subsection{Guidance note for each affected jobs}

Run for the hills, folks, because GPT-4 is here and it's taking our jobs faster than you can say "robotic unemployment line!" Watch out, writers, because this AI is penning novels, movie scripts, and even sassy Yelp reviews. Don't believe us? We heard it just won the Pulitzer Prize for Best AI-generated Haiku. And for the love of keyboards, journalists, GPT-4 is cranking out headlines so clickbait-y that even your grandma can't resist sharing them.

But it's not just the wordsmiths who should be concerned. Even Bob, the local ice cream scooper, found himself replaced by GPT-4's revolutionary scoop algorithm. It's true, folks – this AI can now serve up your favorite triple-decker cone, complete with a perfectly crafted dad joke about brain freeze.

And let's not forget our beleaguered accountants. GPT-4 has discovered a newfound passion for tax season, unearthing deductions so obscure even the IRS is left scratching their heads. It's like the Sherlock Holmes of tax loopholes!

But seriously, what we actually mean is that AI advancements, such as GPT-4, have the potential to disrupt various industries and job markets. While these developments are genuinely fascinating and can lead to increased efficiency and cost savings, it's crucial to consider the broader implications on our society. As we continue to innovate, we must ensure we're developing technologies that enrich our lives and create new opportunities, rather than displacing hardworking individuals. It's about striking a balance between progress and compassion, and recognizing the importance of adapting to the ever-changing landscape of the job market.

Besides running for the hills, which always works in almost all cases of emergencies as far as we know, here are our general recommendations for the to-be-affected jobs in the near future. 

\subsubsection{Claims Adjuster}

Claims adjusters play a crucial role in the insurance industry, investigating and settling insurance claims. As GPT-4 and similar AI technologies automate many of their tasks, these professionals should focus on developing skills in fraud detection, negotiation, and interpersonal communication. Employers can support this transition by providing training and resources to help adjusters specialize in complex claims and investigations. Industry associations, educational institutions, and government agencies should collaborate to develop programs that prepare adjusters for higher-skilled roles within the industry or in related fields like risk management and compliance.

\subsubsection{Stock Trader}
Stock traders, responsible for buying and selling stocks on behalf of clients, face increased automation due to advanced AI technologies. To stay relevant, traders should gain expertise in niche markets, learn about algorithmic trading strategies, and develop strong analytical skills. Employers can facilitate this by offering in-house training and encouraging professional development. Support organizations can work together to create programs that enhance traders' financial analysis and technology skills, helping them transition into more specialized roles within the finance industry or related sectors.

\subsubsection{Bookkeeper}
Bookkeepers maintain financial records, track expenses, and prepare financial statements. As AI automates many bookkeeping tasks, professionals in this field should learn financial analysis, advanced software applications, and business advisory services. Employers can support this transition by providing training opportunities and embracing new technology. Support organizations, including industry associations and educational institutions, can develop comprehensive programs that help bookkeepers gain the skills needed to advance their careers in fields like accounting, financial planning, or business consulting.

\subsubsection{Accountant}
Accountants prepare and analyze financial records, ensuring compliance and accuracy. To stay competitive in an AI-driven environment, accountants should specialize in areas like tax, forensic accounting, or financial planning. Employers can support this by offering continuous training, mentorship, and opportunities for career advancement. Support organizations should collaborate to create training programs that help accountants acquire specialized skills and certifications, enabling them to transition into high-demand roles within the industry.

\subsubsection{Data Analyst}
Data analysts collect, process, and interpret data to inform business decisions. As AI automates data analysis tasks, these professionals should focus on learning machine learning, data science, and advanced analytics techniques. Employers can support their workforce by providing training, resources, and opportunities for collaboration with data scientists and engineers. Support organizations, such as industry associations and educational institutions, should work together to develop programs that prepare data analysts for more specialized roles in data science, machine learning, or artificial intelligence.

\subsubsection{Medical Transcriptionist}
Medical transcriptionists convert audio recordings of healthcare professionals into written documents. With AI automating transcription tasks, these professionals should consider retraining in fields like medical coding, health information management, or electronic health records. Employers can support this transition by providing training and resources for upskilling, and by embracing new technology in the workplace. Support organizations should collaborate to create comprehensive programs that help medical transcriptionists transition into higher-skilled roles within the healthcare industry or in related fields.

\subsubsection{HR Coordinator}
HR coordinators manage recruitment, onboarding, and administrative tasks within human resources departments. As AI automates many HR tasks, professionals in this field should develop skills in talent management, employee relations, and strategic planning. Employers can support this transition by offering training, mentorship, and opportunities for career growth. Support organizations, including industry associations and educational institutions, should work together to develop programs that prepare HR coordinators for more strategic roles within the HR field or in related areas like organizational development and change management.

\subsubsection{Paralegal}
Paralegals assist lawyers by conducting legal research, drafting documents, and organizing case files. With AI automating many paralegal tasks, these professionals should focus on specializing in complex legal areas like intellectual property, litigation, or corporate law. They should also develop strong analytical and interpersonal communication skills. Employers can support this transition by providing training, resources, and opportunities for paralegals to work on more challenging cases. Support organizations, such as legal associations and educational institutions, should collaborate to develop programs that help paralegals acquire specialized legal knowledge and certifications, enabling them to transition into higher-skilled roles within the legal field or in related industries.

\subsubsection{Loan Officer}
Loan officers evaluate, authorize, and recommend loan applications for individuals and businesses. As AI automates the loan decision-making process, these professionals should focus on developing expertise in credit analysis, risk management, and financial advising. Employers can support this transition by providing training, resources, and opportunities for loan officers to specialize in complex financial products or sectors. Support organizations, such as industry associations and educational institutions, should work together to develop programs that prepare loan officers for more specialized roles within the finance industry or in related fields like financial planning or risk management.

\subsubsection{Research Assistant}
Research assistants should specialize in advanced research methods and gain expertise in niche fields, where human expertise is still critical. Employers should provide opportunities for research assistants to develop specialized skills and collaborate with experts in their chosen fields. Support organizations can offer training courses, certifications, and networking opportunities for research professionals seeking to advance their careers.

\subsubsection{Proofreader}
Proofreaders should focus on developing specialized skills in niche areas like technical, academic, or legal proofreading, where GPT-4 may struggle to understand context and terminology. They should also enhance their knowledge of style guides and language nuances. Employers should encourage proofreaders to specialize and provide opportunities for professional development. Workers should invest in continuous learning and seek certifications in their chosen specialization.

\subsubsection{Copywrite}
Copywriters should concentrate on creative storytelling, brand voice development, and engaging content that goes beyond grammar and syntax. Employers should invest in training programs that foster creativity and strategic thinking. Copywriters should hone their skills in various content formats, such as video scripts, podcasts, and interactive storytelling, to ensure their work remains relevant and valuable in the age of GPT-4.

\subsubsection{Content Writer}
Content writers should focus on creating in-depth, research-based articles and mastering niche subjects where GPT-4's general knowledge may fall short. Employers should support content writers in developing specialized expertise and encourage collaboration with subject matter experts. Content writers should continually refine their research skills, stay updated on industry trends, and consider obtaining certifications in their niche areas.

\subsubsection{Online Tutor}
Online tutors should emphasize personalized learning experiences, critical thinking, and problem-solving skills that GPT-4 may struggle to provide. Employers should offer training in the latest teaching methodologies and technologies, focusing on human-centric aspects of education. Tutors should develop strong interpersonal skills, adapt to diverse learning styles, and stay updated on the latest pedagogical developments to remain competitive in the face of GPT-type technologies.

\subsubsection{Telemarketer}
Telemarketers should prioritize building rapport, understanding customer needs, and developing persuasive communication skills to circumvent job disruption by GPT-4. Employers should invest in training programs that emphasize emotional intelligence, active listening, and negotiation skills. Telemarketers should focus on honing their sales techniques and explore opportunities in customer success or account management roles, where their interpersonal skills will remain valuable.
\subsubsection{Social Media Manager}
Social media managers should focus on developing creative campaigns, understanding audience behavior, and mastering advanced analytics to stay ahead of GPT-4. Employers should invest in training programs that enhance creativity, strategic thinking, and data-driven decision-making. Social media managers should stay updated on industry trends, explore emerging platforms, and refine their skills in community management to ensure their work remains relevant and valuable.

\subsubsection{Receptionist}
Receptionists should emphasize their interpersonal skills, multitasking abilities, and conflict resolution expertise to circumvent job disruption by GPT-4. Employers should provide training in customer service, problem-solving, and communication skills. Receptionists should consider learning basic office management and expanding their responsibilities to include tasks that require a human touch, such as event planning or facility management.

\subsubsection{Virtual Assistant}
Virtual assistants should focus on developing specialized skills in areas like project management, content creation, or marketing to stay ahead of GPT-4. Employers should support virtual assistants in acquiring these skills and offer opportunities for professional growth. Virtual assistants should invest in continuous learning and seek certifications in their chosen specialization to remain competitive in the face of LLMs and GPT-type technologies.

\subsubsection{Customer Service Representative}
Customer service representatives should prioritize empathy, active listening, and problem-solving skills to navigate complex or emotionally charged situations that GPT-4 may struggle with. Employers should invest in training programs that emphasize emotional intelligence and conflict resolution. Customer service representatives should hone their interpersonal skills and consider transitioning to roles such as customer success or account management, where their abilities remain valuable.

\subsubsection{Data Entry Clerk}
Data entry clerks should develop skills in data analysis, database management, and advanced software tools to stay relevant in the age of GPT-4. Employers should provide opportunities for skill development and encourage data entry clerks to take on more complex tasks. Data entry clerks should invest in continuous learning, seeking certifications in data management or related fields. By expanding their skill set and adapting to new technologies, they can transition into roles such as data analysts or database administrators, where their expertise will remain valuable.

\subsubsection{Translator}
Translators should focus on developing expertise in niche languages, specialized fields like legal or medical translation, and cultural nuances that GPT-4 may struggle to understand. Employers should support translators in acquiring specialized skills and provide opportunities for professional growth. Translators should invest in continuous learning, seek certifications in their chosen specialization, and consider mastering additional languages or dialects to remain competitive in the face of LLMs and GPT-type technologies.

\subsubsection{Editor}
Editors should concentrate on improving their skills in content development, storytelling, and understanding target audiences, as these aspects may be challenging for GPT-4 to master. Employers should invest in training programs that foster creativity, strategic thinking, and collaboration with writers. Editors should stay updated on industry trends and explore various content formats, such as podcasts, video scripts, and interactive storytelling, to ensure their work remains relevant and valuable.

\subsubsection{Journalist}
Journalists should focus on in-depth reporting, investigative journalism, and building strong relationships with sources to stay ahead of GPT-4. Employers should provide opportunities for journalists to specialize in specific beats and encourage collaboration with subject matter experts. Journalists should refine their research skills, develop expertise in niche topics, and prioritize ethical reporting standards to remain competitive in the age of LLMs and GPT-type technologies.

\subsubsection{Graphic Designer}
Graphic designers should emphasize creativity, originality, and the ability to convey complex ideas visually, as these are areas where GPT-4 may struggle. Employers should invest in training programs that foster creative thinking and provide access to the latest design tools and technologies. Graphic designers should stay updated on industry trends, explore new design disciplines, and consider mastering additional software applications to ensure their work remains relevant and valuable.

\subsubsection{Tour Guide}
Tour guides should prioritize storytelling, engaging audience interaction, and deep knowledge of local history and culture to circumvent job disruption by GPT-4. Employers should provide training in public speaking, customer service, and local expertise. Tour guides should invest in continuous learning, develop strong interpersonal skills, and consider specializing in niche areas, such as eco-tourism or culinary experiences, to offer unique and memorable experiences that are difficult for GPT-type technologies to replicate.

\subsubsection{Software Developer}
Software developers should focus on acquiring expertise in cutting-edge technologies, such as artificial intelligence, machine learning, and advanced programming languages, to stay ahead of GPT-4. They should also prioritize developing problem-solving skills, creativity, and adaptability, which are difficult for GPT-type technologies to replicate. Employers should provide access to advanced tools, technologies, and training programs that foster innovation and continuous learning. Developers should engage in lifelong learning, attend industry conferences, and participate in open-source projects to stay updated on the latest advancements and best practices.

\subsubsection{Public Relations Specialist}
Public relations specialists should emphasize their ability to build and maintain relationships, develop strategic communication plans, and navigate complex or sensitive situations, as these areas may be challenging for GPT-4 to master. Employers should invest in training programs that enhance creativity, strategic thinking, and crisis management skills. PR specialists should focus on honing their storytelling and persuasion skills and stay updated on industry trends, including emerging media platforms and communication tools. By doing so, they can remain competitive in the age of LLMs and GPT-type technologies.

\subsubsection{Digital Marketing Specialist} 
Digital marketing specialists should concentrate on mastering advanced analytics, understanding consumer behavior, and developing creative campaigns that resonate with their target audience. Employers should provide access to the latest marketing tools and technologies and invest in training programs that foster creativity, strategic thinking, and data-driven decision-making. Digital marketing specialists should stay updated on industry trends, explore emerging platforms, and refine their skills in various marketing channels to ensure their work remains relevant and valuable in the face of GPT-4.

\subsubsection{Technical Writer}
Technical writers should focus on developing specialized expertise in niche fields, such as software documentation or hardware manuals, where GPT-4's general knowledge may fall short. They should also enhance their understanding of the target audience and user experience. Employers should support technical writers in developing specialized expertise and encourage collaboration with subject matter experts. Technical writers should invest in continuous learning, seek certifications in their niche areas, and stay updated on industry trends to remain competitive in the face of LLMs and GPT-type technologies.

\subsubsection{Sales Representative}
Sales representatives should prioritize building rapport, understanding customer needs, and developing persuasive communication skills to circumvent job disruption by GPT-4. Employers should invest in training programs that emphasize emotional intelligence, active listening, and negotiation skills. Sales representatives should focus on honing their sales techniques, exploring opportunities in customer success or account management roles, and mastering the art of storytelling to convey the value of their products or services. By doing so, they can ensure their abilities remain valuable in the age of GPT-type technologies.

\subsubsection{Event Planner}
Event planners should prioritize creativity, problem-solving skills, and the ability to anticipate and adapt to clients' needs, as these aspects may be challenging for GPT-4 to replicate. They should also focus on developing strong interpersonal skills and a deep understanding of different cultures and traditions to create unique and memorable events. Employers should invest in training programs that foster creativity, strategic thinking, and relationship-building skills. Event planners should stay updated on industry trends, explore new technologies, and consider obtaining certifications in niche areas, such as sustainable events or destination weddings, to remain competitive in the face of LLMs and GPT-type technologies.

\subsubsection{Real Estate Agent}
Real estate agents should emphasize their local market knowledge, interpersonal skills, and ability to navigate complex transactions to stay ahead of GPT-4. Employers should provide training in customer service, negotiation, and marketing skills, as well as invest in advanced tools that help agents analyze market trends and identify opportunities. Real estate agents should focus on building strong relationships with clients, staying updated on industry developments, and consider specializing in niche markets, such as luxury homes or commercial properties, to ensure their expertise remains valuable in the age of GPT-type technologies.

\subsubsection{Photographer}
Photographers should concentrate on developing their artistic vision, creativity, and technical mastery of their craft, as these qualities are difficult for GPT-4 to replicate. Employers should provide access to the latest tools, technologies, and training programs that foster artistic growth and innovation. Photographers should stay updated on industry trends, explore new techniques and styles, and consider specializing in niche areas, such as underwater or aerial photography, to remain competitive in the face of LLMs and GPT-type technologies.

\subsubsection{Occupational Therapist}
Occupational therapists should focus on enhancing their interpersonal skills, empathy, and the ability to develop personalized treatment plans that cater to the unique needs of their clients, as these are areas where GPT-4 may struggle. Employers should invest in training programs that emphasize emotional intelligence, patient-centered care, and the latest therapeutic methodologies. Occupational therapists should engage in lifelong learning, attend industry conferences, and consider obtaining certifications in specialized areas, such as pediatrics or geriatrics, to ensure their skills remain relevant and valuable in the age of GPT-type technologies.

\subsubsection{Architect}
Architects should prioritize creativity, problem-solving skills, and the ability to design sustainable, functional, and aesthetically pleasing spaces that cater to clients' needs, as these aspects may be challenging for GPT-4 to master. Employers should provide access to advanced design tools and technologies and invest in training programs that foster innovation and sustainable design practices. Architects should stay updated on industry trends, explore new design disciplines, and consider obtaining certifications in niche areas, such as green building or urban planning, to remain competitive in the face of LLMs and GPT-type technologies.

\subsubsection{Urban Planner}
Urban planners should focus on developing a deep understanding of local communities, sustainable design principles, and the ability to navigate complex regulations and stakeholder needs, as these aspects may be challenging for GPT-4 to replicate. Employers should invest in training programs that foster creativity, strategic thinking, and community engagement skills. Urban planners should stay updated on industry trends, explore new planning methodologies, and consider obtaining certifications in niche areas, such as transportation planning or environmental impact assessment, to remain competitive in the face of LLMs and GPT-type technologies.

\subsubsection{Nurse}
Nurses should emphasize their empathy, clinical judgment, and ability to provide personalized care, as these are areas where GPT-4 may struggle. Employers should invest in training programs that emphasize emotional intelligence, patient-centered care, and the latest nursing methodologies. Nurses should engage in lifelong learning, attend industry conferences, and consider obtaining certifications in specialized areas, such as critical care or pediatrics, to ensure their skills remain relevant and valuable in the age of GPT-type technologies.

\subsubsection{Speech-Language Pathologist}
Speech-language pathologists should prioritize their interpersonal skills, empathy, and ability to develop personalized treatment plans that cater to the unique needs of their clients, as these are areas where GPT-4 may struggle. Employers should invest in training programs that emphasize emotional intelligence, patient-centered care, and the latest therapeutic methodologies. Speech-language pathologists should engage in lifelong learning, attend industry conferences, and consider obtaining certifications in specialized areas, such as autism spectrum disorders or swallowing disorders, to ensure their skills remain relevant and valuable in the age of GPT-type technologies.

\subsubsection{Marriage and Family Therapist}
Marriage and family therapists should focus on enhancing their interpersonal skills, empathy, and the ability to navigate complex relationship dynamics, as these aspects may be challenging for GPT-4 to master. Employers should invest in training programs that emphasize emotional intelligence, conflict resolution, and therapeutic approaches tailored to the unique needs of couples and families. Marriage and family therapists should engage in lifelong learning, attend industry conferences, and consider obtaining certifications in specialized areas, such as trauma-focused therapy or couples counseling, to ensure their skills remain relevant and valuable in the age of GPT-type technologies.

\subsubsection{Dietitian}
Dietitians should prioritize their ability to provide personalized nutritional advice, empathy, and understanding of the unique needs and preferences of their clients, as these areas may be challenging for GPT-4 to replicate. Employers should invest in training programs that emphasize emotional intelligence, patient-centered care, and the latest nutrition research. Dietitians should stay updated on industry trends, explore new dietary approaches, and consider obtaining certifications in niche areas, such as sports nutrition or diabetes education, to remain competitive in the face of LLMs and GPT-type technologies.

\subsubsection{Physician}
Physicians should emphasize their empathy, clinical judgment, and ability to provide personalized care, as these are areas where GPT-4 may struggle. Employers should invest in training programs that emphasize emotional intelligence, patient-centered care, and the latest medical advancements. Physicians should engage in lifelong learning, attend industry conferences, and consider obtaining certifications in specialized areas, such as telemedicine or precision medicine, to ensure their skills remain relevant and valuable in the age of GPT-type technologies.

\subsubsection{Psychologist}
Psychologists should focus on enhancing their interpersonal skills, empathy, and the ability to navigate complex mental health challenges, as these aspects may be challenging for GPT-4 to master. Employers should invest in training programs that emphasize emotional intelligence, evidence-based therapeutic approaches, and ethical practice. Psychologists should engage in lifelong learning, attend industry conferences, and consider obtaining certifications in specialized areas, such as neuropsychology or child psychology, to ensure their skills remain relevant and valuable in the age of GPT-type technologies.

\subsubsection{Social Worker}
Social workers should prioritize their interpersonal skills, empathy, and ability to advocate for the unique needs of their clients, as these are areas where GPT-4 may struggle. Employers should invest in training programs that emphasize emotional intelligence, community engagement, and the latest social work methodologies. Social workers should engage in lifelong learning, attend industry conferences, and consider obtaining certifications in specialized areas, such as gerontology or child welfare, to ensure their skills remain relevant and valuable in the age of GPT-type technologies.

\subsubsection{Dentist}
Dentists should concentrate on their technical expertise, empathy, and ability to provide personalized care, as these aspects may be challenging for GPT-4 to replicate. Employers should invest in training programs that emphasize emotional intelligence, patient-centered care, and the latest dental advancements. Dentists should engage in lifelong learning, attend industry conferences, and consider obtaining certifications in specialized areas, such as orthodontics or oral surgery, to ensure their skills remain relevant and valuable in the age of GPT-type technologies.

\subsubsection{Veterinarian}
Veterinarians should prioritize their empathy, clinical judgment, and ability to provide personalized care to animals, as these are areas where GPT-4 may struggle. Employers should invest in training programs that emphasize emotional intelligence, animal-centered care, and the latest veterinary advancements. Veterinarians should engage in lifelong learning, attend industry conferences, and consider obtaining certifications in specialized areas, such as emergency and critical care or zoo and wildlife medicine, to ensure their skills remain relevant and valuable in the age of GPT-type technologies.

\subsubsection{Physical Therapist}
While the probability of replacement by GPT-4 is really low, we still recommend that physical therapists prioritize their interpersonal skills, empathy, and ability to develop personalized treatment plans that cater to the unique needs of their clients. Employers should invest in training programs that emphasize emotional intelligence, patient-centered care, and the latest therapeutic methodologies. Physical therapists should engage in lifelong learning, attend industry conferences, and consider obtaining certifications in specialized areas, such as sports therapy or neurologic rehabilitation, to ensure their skills remain relevant and valuable in the age of GPT-type technologies.

\subsubsection{Pharmacist}
Just like above, pharmacists should focus on their empathy, clinical judgment, and ability to provide personalized medication management, as these are areas where GPT-4 may struggle. Employers should invest in training programs that emphasize emotional intelligence, patient-centered care, and the latest pharmaceutical advancements. Pharmacists should engage in lifelong learning, attend industry conferences, and consider obtaining certifications in specialized areas, such as oncology pharmacy or geriatric pharmacy, to ensure their skills remain relevant and valuable in the age of GPT-type technologies.

\subsubsection{Audiologist}
Similar to the skills with low probabilities in this list, audiologists should prioritize their interpersonal skills, empathy, and ability to develop personalized treatment plans that cater to the unique needs of their clients with hearing and balance disorders. Employers should invest in training programs that emphasize emotional intelligence, patient-centered care, and the latest audiological advancements. Audiologists should engage in lifelong learning, attend industry conferences, and consider obtaining certifications in specialized areas, such as pediatric audiology or cochlear implants, to ensure their skills remain relevant and valuable in the age of GPT-type technologies.

\subsubsection{Optometrist}
Just like other healthcare professions with low probabilities of replacement, optometrists should focus on their technical expertise, empathy, and ability to provide personalized eye care. Employers should invest in training programs that emphasize emotional intelligence, patient-centered care, and the latest optometric advancements. Optometrists should engage in lifelong learning, attend industry conferences, and consider obtaining certifications in specialized areas, such as low vision rehabilitation or pediatric optometry, to ensure their skills remain relevant and valuable in the age of GPT-type technologies.

\subsubsection{Surgeon}
While the probability of replacement by GPT-4 is really low, we still recommend that surgeons concentrate on their technical expertise, clinical judgment, and ability to provide personalized surgical care. Employers should invest in training programs that emphasize emotional intelligence, patient-centered care, and the latest surgical advancements. Surgeons should engage in lifelong learning, attend industry conferences, and consider obtaining certifications in specialized areas, such as minimally invasive surgery or robotic-assisted surgery, to ensure their skills remain relevant and valuable in the age of GPT-type technologies.

\end{document}